\documentclass[twocolumn,english,aps,pre,showpacs]{revtex4}
\usepackage[T1]{fontenc}
\usepackage[latin1]{inputenc}
\usepackage{float}
\usepackage{amsmath}
\usepackage{graphicx}
\usepackage{amssymb}

\makeatletter

\usepackage{babel}
\makeatother
\begin{document}

\preprint{preprint}

\title{Phonon modes in the Frenkel-Kontorova chain:\\
exponential localization and the number theory properties of frequency bands}

\author{O.V. Zhirov}
\email{zhirov@inp.nsk.su}
\affiliation{Budker Institute of Nuclear Physics, 630090 Novosibirsk, Russia}

\author{G. Casati}
\email{giulio.casati@uninsubria.it}
\affiliation{International Center for the Study of Dynamical 
Systems, Universit\`a degli Studi dell'Insubria and\\
Istituto Nazionale per la Fisica della Materia, 
Unit\`a di Como, Via Valleggio 11, 22100 Como, Italy and\\
Istituto Nazionale di Fisica Nucleare, 
Sezione di Milano, Via Celoria 16, 20133 Milano, Italy}

\author{D.L. Shepelyansky}
\email{dima@irsamc.ups-tlse.fr}
\affiliation{Laboratoire de Physique Quantique, UMR 5626 du CNRS, Universit\'{e}
Paul Sabatier, 31062 Toulouse, France}

\date{\today{}}

\begin{abstract}
We study numerically phonon modes of the classical one-dimensional
Frenkel-Kontorova chain, in the regime of pinned phase characterized 
by the phonon gap and devil's staircase, as well as by a large number
of states (configurational excitations), which energy splitting from the ground
state is exponentially small. We demonstrate,  these states behave like 
disorder media: their phonon modes are {\it exponentially} localized, 
in contrast to the phonon modes in the ground state, where phonons are 
{\it prelocalized} only.  

We demonstrate also, the phonon frequency spectrum of the ground state has an 
hierarchical structure, a direct manifestation of hierarchical spatial structure, 
found for the ground state of the FK chain in our recent work. 
\end{abstract}
\maketitle

\section{Introduction}

The most trivial disorder originates in media due to random static 
impurities (see, e.g. \cite{Pastur}). However, another but very interesting 
possibilities are glasses, which have a huge number of (meta)stable 
degenerated states. Originally glassy system has
a homogeneous Hamiltonian with no intrinsic random parameter, and disorder
occurs in it dynamically. Recently \cite{Zh01} we have demonstrated that a popular 
Frenkel-Kontorova model \cite{FK38} presents an example of glassy system,
which has a lot of static states, known as {\it configurational excitations}
of the classical ground state, with energy splitting extremely (exponentially)
small. As it was shown in \cite{Zh03}, this model has a nontrivial quantum
dynamics, the quantum phase transition: if quantum parameter exceeds some 
critical value, the "pinned" glassy phase turns into "sliding" phonon gas.

The Frenkel-Kontorova model (FK) \cite{FK38} is widely used 
\cite{Naba67,Pokr84,Ying71,Piet81,Au78,Au83a,Flor96,Weis96,Brau97,Cons00}
in the solid state physics to get insight on generic properties of
noncommesurate systems. Its ground state, which is rather quasiperiodical
\cite{Au78,Au83b,Zh01} than periodical, attracts also an attention
\cite{Au78,Au83a,Au90,Burk96,Hu00} as some interplay \cite{Burk96}
between order and disorder \cite{Sin77}.

This model describes a chain of atoms/particles interacting with elastic
forces, placed in periodic potential, which period differs from a
mean interparticle distance. The ground state (GS) of this model is
defined as a static, equilibrium configuration of the chain, that
corresponds to the {\it absolute} minimum of the chain potential
energy. The ground state is unique and has some special order of particles
in the chain, that was discovered by Aubry \cite{Au78,Au83b} more
than twenty years ago. The positions of atoms in the chain are described
by an area preserving map, which is well known in the field of dynamical
chaos as the Chirikov standard map \cite{Chir79}. The ratio of the
mean interparticle distance to a period of the external potential
in the FK model determines the rotation number of the invariant curves
of the map, while the amplitude of the periodic potential gives the
value of the dimensionless parameter $K$. For $ K<K_{c} $
the KAM curves are smooth and the spectrum of longwave phonon excitations
in the chain is characterized by a linear dispersion law starting
from zero frequency. In this regime the chain can freely slide along
the external field (the ``sliding'' phase). On the contrary, for
$ K>K_{c} $ the KAM curves are destroyed and replaced by an invariant
Cantor set, which is called cantorus. In this regime the phonon spectrum
has a gap, and the chain is pinned (``pinned'' phase). Later,
on the example of Ising spin model to which the FK model can be approximately
reduced \cite{Be80} it has been shown \cite{Vall86} that the GS
has some well defined hierarchical structure, which particular detailes
are determined by number properties of the ratio of the mean interparticle
distance to the period of the external field. Recently our numerical 
study \cite{Zh01} of the original FK model in the pinned phase has shown, 
that the GS has indeed an hierarchical structure, but in some important 
detailes different from predicted in \cite{Vall86}. 

In short, we put our attention to the striking fact, that
in the pinned phase of the FK chain there are some particles, which positions
exponentially close to bottoms of wells of the external potential:
corresponding external force acting to such particle is extremely close to zero.
Obviously, in static equilibrium each of these particles can be considered as some 
dummy ``glue'' that only couples two adjacent parts of the chain, 
which ends has almost identical (exponentially close) tension forces. 

Another important observation is that small deviations of glue particles from 
their well bottoms are {\it groupped} 
into well defined hierarchically ordered scales. Now, if one cut the
chain into fragments via glue particles, which belong to level with
least deviation from well bottoms, one gets several fragments of {\it two}
sizes, or two species of some ``bricks''\footnote{%
The number of brick species depends on the number properties of the
ratio of a mean interparticle distance to a period of the external
field, the chain ``rotation number'' $\nu $. The most simple picture
quoted here assumes that this rotational number approximates the golden
mean value. 
}. Then one may repeat the procedure, cutting ``bricks'' via ``glue''
particles that belong to next scale of deviations and getting two
new species of smaller bricks, and so far. At the last step one gets
two species of smallest possible bricks with {\it no} glue particle
inside: smaller bricks $ A^{(0)} $ which consist of 2 particles
inside a single well, and larger brick $ B^{(0)} $ which consist
of 4 particles (two pairs in two adjacent wells). In this way one
gets an hierarchically ordered set of brick species 
$ \left\{ A^{(i)},B^{(i)}\right\} $
with very simple composition rules \cite{Zh01}:

\begin{eqnarray}
A^{(i+1)} & = & B^{(i)}gA^{(i)}gB^{(i)},\label{cr1} \\
B^{(i+1)} & = & B^{(i)}gA^{(i)}gB^{(i)}gA^{(i)}gB^{(i)},\label{cr2} 
\end{eqnarray}

where symbol $ g $ denotes of an insertion of glue particle, which
``glues'' two adjacent bricks. The difference of tension forces
at boundaries of bricks $ A^{(i)} $ and $ B^{(i)} $ is exponentially
small and decrease rapidly with a number of the hierarcy level. In
principle, these rules are sufficient to construct a GS for a FK chain
of any length, if the rotation number parameter of the chain approximates
the mean golden value $ \nu =(\sqrt{5}-1)/2 $. 

Besides the GS there exists  ``configurational excitation states''
(CES), presented by static equilibrium configurations corresponding
to {\it local} (rather than absolute) minima of the chain potential, 
with energy very close to GS. Within the picture just outlined above 
CES correspond to different permutations of bricks \cite{Zh01}, 
therefore the number of them can be combinatorically huge. 
At any accessible small temperature their contributions can dominate 
over the contribution of GS.

In this paper we address to phonon excitations of the chain, small
vibrations around static GS and CES configurations of the chain. These
excitations are relevant as for heat transport properties \cite{Tong99,Hu00}
of the chain, as for some quantum effects \cite{Hu01}, especially
in the quasiclassical limit. As in the previous paper, we concentarte
on the case of pinned phase of the chain, which corresponds to a nonzero
phonon gap. We start with analysis of the structure of the phonon
frequency spectrum in the GS. It is well known, that this spectrum
is splitted into bands \cite{Au83a,Pe83} but, to our knowledge, up
to now there is no clear explanation for origin of its splitting into
particular bands and subbands. We show, that this splitting is a direct
consequence of particular spatial structure of the chain in its GS.
We have found, that this structure is also hierarchically ordered,
with definite resemblance and distinctions with respect to a spatial
structure of underlying GS.

Localisation properties of phonons in incommensurate one-dimentional
chains are intesively studied in recent works \cite{Burk96,Keto97,Tong99,Hu00},
with strong indications \cite{Keto97,Tong99}, that phonon modes in
the GS of FK chain are not localized, and even at edges of frequency
bands they are rather {\it prelocalized}, than localized. We study
also phonon properties as in GS, as in CES of the chain. Our results
confirm, that in the GS phonon modes are only prelocalized. However,
the situation appears to be quite different for CES. Even for CES,
which energy splitting (in natural problem scale) $ \Delta U\leq 10^{-12} $
there are phonon modes, which are localized exponentially. Moreover,
for CES with higher splitting \( \Delta U \) we see, that there are
entire bands of exponentially localized phonon modes.

\section{The model.}

The Hamiltonian of the FK model is 

\begin{equation}
\label{HFK}
H=\sum _{i=1}^{s}\frac{P_{i}^{2}}{2}+\frac{(x_{i}-x_{i-1})^{2}}{2}-K\cos x_{i}.
\end{equation}

The first term in the Hamiltonian is a kinetic energy, where we put
masses of particles $ m=1 $, the second term describes interparticle
interaction with elasticity coefficient put to unity, while the third
term corresponds to particle interaction with external periodical
field with coupling constant $ K $. All $ s $ particles are
distributed over $ r $ period/wells of the external potential,
which period, without any loss of generality is taken equal to $ 2\pi $.
The ratio $ \nu =r/s $ gives \cite{Au78} the rotational number
of corresponding standard map \cite{Chir79}. 

We assume periodical boundary conditions: $ P_{0}\equiv P_{s} $,
$ x_{0}\equiv x_{s}-L $, where $ L=2\pi r $ is the length of
the chain. In our subsequent analysis we take (as some {\it typical}
example of FK chain) the chain with $ r/s=377/610 $ as an approximation
of the golden mean value $ \bar{\nu }=(\sqrt{5}-1)/2 $, and parameter
$ K=2 $ well above the critical value 
$ K_{c}(\nu =\bar{\nu })=0.971\ldots  $\cite{Gree79}.
The technique to obtain GS and CES is described in our previous work
\cite{Zh01}. Let us stress, that validity of the GS can be checked
either by direct analysis \cite{Zh01} of its spatial structure, or
proved by monotonous behavior \cite{Au83b} of its hall function. 
Note, that the GS is almost degenerate with enormous number of CES: there are
hundreds of states with $ \Delta U=U_{ECS}-U_{GS}\leq 10^{-80} $,
while the number of ECS with $ \Delta U\leq 10^{-12} $ exceeds
$ 10^{9} $!

Effective Hamiltonian for phonon modes can be obtained by expansion
up to second order terms of the original FK chain Hamiltonian (\ref{HFK})
around a chosen static configuration of the chain:
\begin{equation}
\label{Hph}
H^{ph}=\sum ^{s}_{i=1}\frac{\Pi _{i}^{2}}{2}
       +\frac{1}{2}\sum ^{s}_{i,k=1}\frac{\partial U}{\partial x_{i}\partial x_{k}}\psi _{i}\psi _{k},
\end{equation}
where $ \psi _{i}=x_{i}-\bar{x}_{i} $ are small deviations of particles
from their equilibrium static positions $ \bar{x}_{i} $, and $ \Pi _{i} $
are corresponding particle momenta. The elasticity matrix
\begin{equation}
\label{Rik}
   R_{ik}\equiv \frac{\partial ^{2}U}{\partial x_{i}\partial x_{k}}=
                K\cos (\bar{x}_{i})\delta _{ik}-\delta _{i,k+1}-\delta _{i+1,k}.
\end{equation}
Solving the eigenvalue problem
\begin{equation}
\label{EigVal}
\left( R-\omega ^{2}I\right) \psi =0,
\end{equation}
numerically, we get both the spectrum of phonon frequencies and corresponding
vectors of phonon modes.

\section{Phonon frequency spectrum}

\label{sPhFr}

It is well known for a long time \cite{Au83a,Pe83}, that frequency
spectrum of phonons in the Frenkel-Kontorova chain is splitted into
several bands. Main features of the spectrum for a chain in the GS
look as very universal: (i) the number of main bands are independent
of the parameter $ K $ as well as of a length of the chain; 
(ii) it was also noticed in \cite{Burk96} (without any explanation, either)
that boundaries of the bands correspond to Fibonacci numbers; 
(iii) replacement in interparticle interactions elastic forces by the Lenard-Jones
potential does not change qualitatively a pattern of splitting phonon
spectrum into band \cite{Burk96}. 

Now we can explain these and more features as direct consequence of
particular spatial structure \cite{Zh01} of the GS, which main relevant
detailes are summarized in Introduction. The key point of our explanation
is that the GS consists of large number of almost identical elements.
For clarity sake, we consider a particular case of the FK chain with
$ r/s=377/610 $, but all our arguments can be easily generalizied
for a chain of different size.

Applying the picture of CS spatial structure outlined in Introduction
to our particular case, we can see, that our CS can be presented as
a composition of two bricks of 3d level of hierarcy:
\begin{equation}
\label{xL3}
610=(g232g376),
\end{equation}
with subsequent expansion, in accordance with composition rules (\ref{cr1}),(\ref{cr2}),
into bricks of 2nd level:
\begin{equation}
\label{xL2}
232=(88g54g88),\quad 376=(88g54g88g54g88),
\end{equation}
which, in turn, are expanded into bricks of 1st level:
\begin{equation}
\label{xL1}
54=(20g12g20),\quad 88=(20g12g20g12g20),
\end{equation}
that, at last, are expanded into basic bricks of zero level:
\begin{equation}
\label{xL0}
12=(4g2g4),\quad 20=(4g2g4g2g4).
\end{equation}
In this way it is easy to calculate, that the chain can be cut into
pieces either as 
\begin{equation}
\label{xS0}
144\times g+55\times (2)+89\times (4),
\end{equation}
 or
\begin{equation}
\label{xS1}
34\times g+13\times (12)+21\times (20),
\end{equation}
 or
\begin{equation}
\label{xS2}
8\times g+3\times 54+5\times 88,
\end{equation}
or \( 2\times g+(232)+(376) \), see (\ref{xL3}). Mutual static disturbance
of bricks is small and decreases exponentially with a level of hierarcy,
therefore we have at any level of hierarcy a sequence in some order
of two species of almost identical structures (bricks).

Now let us make some important remark. In our study in \cite{Zh01}
of {\it static} structure of the chain, the glue particles play a
some passive role only, which results in their specification as some
dummy ``glue''. However, in {\it dynamical}
problem of motion particles in the chain, that we now address to, 
the role of glue particles becomes more important,  since they have
masses. Actually, in this case one should consider
as repeating structures {\it elementary cells}, which we get if attach
to each brick (e. g. at the left side) one glue particle\footnote{%
Note, that at any level of hierarcy a number of glue particles is
equal to the total number of bricks.
}.

\begin{figure}
{\centering \resizebox*{0.9\columnwidth}{!}{\rotatebox{90}{\includegraphics{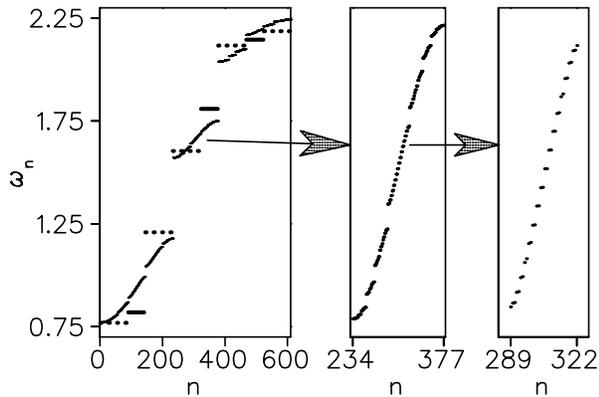}}} 
\par}
\caption{\label{FrGS}Frequency spectrum for FK model in the ground state.
In the left box:the total spectrum of chain, in the middle and right
boxes the central band and its central subband are shown with greater
resolution. Solid and dashed horizontal lines show eigenfrequencies
for isolated cells 
\protect\( \alpha ^{(0)}\protect \) and \protect\( \beta ^{(0)}\protect \)
respectively. }
\end{figure}

At the basic level of hierarcy one have 55 elementary cells of 3 particles
and 89 elementary cells of 5 particles, let them be called as cells
\( \alpha ^{(0)}=(g2) \) and \( \beta ^{(0)}=(g4) \), respectively.
Each elementary cell, being isolated (and, e.g. periodically closed)
has its own eigenfrequencies: three for a cell \( \alpha ^{(0)} \)
and five for a cell \( \beta ^{(0)} \). Then, if these cells in the
chain be really isolated, then, see Fig.\ref{FrGS} we get three bands
each of 55 degenerated states (solid lines) which belongs to cells
\( \alpha ^{(0)} \), and five bands each of 89 degenerated states
(dotted lines).

Due to cells interactions all degenerated frequencies are really splitted
into finite width bands, see Fig.\ref{FrGS}. The closer inspection
of frequency spectrum shows, that seven largest breaks cut the exact
spectrum in bands $ n=(1,89) $, $ (90,144) $, $ (145,233) $,
$ (234,322) $, $ (323,377) $, $ (378,467) $, $ (468,523) $,
$ (524,610) $ so the band widths have the same order 89, 55, 89,
89, 55, 89, 55, 89, exactly as the sequence of solid and dotted horizontal
lines in Fig.\ref{FrGS}!

Now, let us remind, that the sequence of cells $ \alpha ^{(0)} $
and $ \beta ^{(0)} $ in our GS is not random but they belong to
34 cells of the next hierarcy level, see (\ref{xS1}): 13 cells $ \alpha ^{(1)}=(g12)=(g4g2g4) $
and 21 cells $ \beta ^{(1)}=(g20)=(g4g2g4g2g4) $. Again, if this
cells would be decoupled, we could see bands, that consist of 13 and
21 degenerated states. Indeed, in the middle part of the Fig.\ref{FrGS},
where the central band is shown with greater resolution, we see a
sequence of bands, that contain 21,13, 21, (21+13), 21, 13, 21 number
of states, which is similar to that we have at the main level of hierarcy!\footnote{%
Here two central subbands of widths 21 and 13 are merged due to their
overlapping. 
} However, at higher resolution (the right box in Fig.\ref{FrGS})
the spectrum in central band becomes structureless.

This is not surprising, since the ``cells'' we have introduced
are not weakly coupled objects with respect to phonon modes. In fact,
what cells provides, is that some chain fragments have fixed periods;
these fragments form locally their band structure. If bands of different
fragments overlap, their levels are collectivised into one common
band. The central band is the case, however in this case band levels
are less sensitive to small variations introduced by extra regularities
of next levels of hierarcy of spatial chain structure.

However, a quite new interesting phenomenon occurs, if one consider
band, which belongs to a cell $ \beta ^{(0)} $ but does {\it not}
overlap with any band of the cell $ \alpha ^{(0)} $. Phonons with
frequencies inside this band will be damped along the cells of the
kind $ \alpha ^{(0)} $, these cells will play a role of some 
``potential barriers'', which decouple cells of the kind $ \beta ^{(0)} $
each from other. Our chain contains 89 cells \( \beta ^{(0)} \) and
55 cells $ \alpha ^{(0)} $, which is like to 89 particles separated
by 55 barriers, or distributed among 55 wells. Now we have got a new
effective Frenkel-Kontorova chain, where cells $ \beta ^{(0)} $
play role some particles with potential barriers $ \alpha ^{(0)} $
among them. For a clarity sake, let us take more graphical notations
for these effective ``barrier'' $ \wedge \, =\alpha ^{(0)} $,
and ``particle'' $ \bullet \, =\beta ^{(0)} $. Then, seconary
bricks, which occur in this effective FK chain, are 
$ \widetilde{A}=\wedge \bullet \bullet \wedge  $
and $ \widetilde{B}=\wedge \bullet \bullet \wedge \bullet \bullet \wedge  $,
and corresponding elementary cells can be obtained adding to bricks
one ``glue'' particle at the left: 
$ \widetilde{\alpha }=\bullet \wedge \bullet \bullet \wedge  $,
$ \widetilde{\beta }=\bullet \wedge \bullet \bullet \wedge \bullet \bullet \wedge  $,
having again three and five eigenfrequencies, respectively. Our effective
chain has 8 cells $ \widetilde{\alpha } $ and 13 cells $ \widetilde{\beta } $.
Now, if in the frequency spectrum of new {\it effective} FK chain
we take a band, which belongs to the cell $ \widetilde{\beta } $
but not to the cell $ \widetilde{\alpha } $, it must contain 13
states.

\begin{figure}
{\centering \resizebox*{0.9\columnwidth}{!}{\rotatebox{90}{\includegraphics{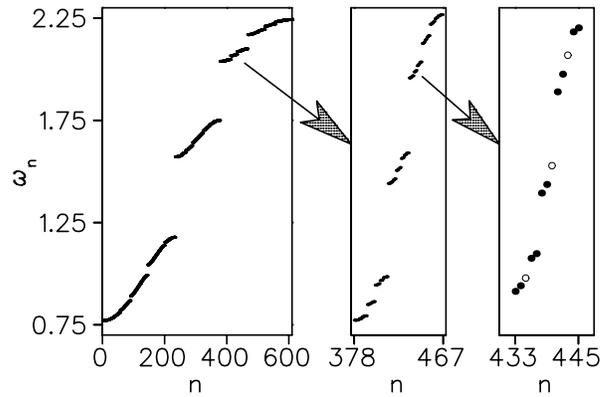}}} \par}

\caption{\label{FrGS4}The same, as in Fig.\ref{FrGS}, but here a {\it fourth}
band and its {\it fourth} subband are shown with greater resolution.
In right figure box open and closed circles correspond to phonon modes
of the cells \protect\( \widetilde{\widetilde{\alpha }}\protect \)
and \protect\( \widetilde{\widetilde{\beta }}\protect \) , respectively.}
\end{figure}

We can repeat the whole procedure ones more; then we come to next
level effective Frenkel-Kontorova chain with 13 effective particles
$ \widetilde{\beta } $ distributed among 8 potential barriers $ \widetilde{\alpha } $,
and to next generation cells, one of the kind $ \widetilde{\widetilde{\alpha }} $
and two of the kind $ \widetilde{\widetilde{\beta }} $. Numerical
data presented in Fig.\ref{FrGS4} confirm our picture in all the
detailes. Here we consider the fourth band of spectra, which is well
resolved from other bands, see, the left box of Fig.\ref{FrGS4}.
It contains 89 states, the number of cells $ \beta ^{(0)} $. In
the middle box we take again the fourth band which in turn contains
13 states, the number of cells $ \widetilde{\beta } $. At last,
in the right box we show by open circles 3 states of the cell 
$ \widetilde{\widetilde{\alpha }} $
and by closed ones $ 2\times 5 $ states of two cells $ \widetilde{\widetilde{\beta }} $.

Note, that this new kind of hierarcy is quite different from that
we found in \cite{Zh01} with respect to spatial structure, since
the transformation rules between levels of hierarcy are more complicated.

\begin{figure}
{\centering \resizebox*{0.9\columnwidth}{!}{\rotatebox{90}{\includegraphics{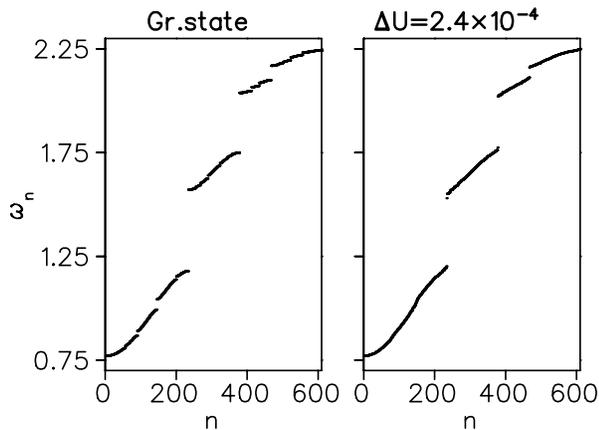}}} \par}

\caption{\label{FrGS_EX}Comparison of the ground state (left box) and excited
state (rigth box) frequency spectra.}
\end{figure}

In conclusion of this section it should be stressed, that all the universal features
of the global band structure mentioned at the beginning are goverened
by nearest order in the chain. In contrast, the fine structure depends
crucially on the far order.. The latter is destroyed in the CES, therefore
in CES the fine structure is washed out, and frequency spectra become
smoother, see Fig.\ref{FrGS_EX}. In particular, we see also in Fig.\ref{FrGS_EX}
how two lowest bands are merged into common one.

\section{Phonon localization}

\label{sPhLoc}

The ordering of bricks, which persists in the GS, is gradually destroyed
with configurational excitation of the chain. The lowest excitations
are equivalent destruction of largest bricks due to permutations of
bricks of preceeding level of hierarcy: the smaller bricks are destroyed,
the higher is the excitation energy \cite{Zh01}. The number of almost
degenerate CES, originating from different bricks permutation is combunatorically
large, therefore some arbitrary chosen CES has rather {\it random}
sequence of bricks, and is similar to disordered media, e.g. spin
glass system.

From the very beginning it is clear, that perfect randomization of
bricks order may result in Anderson localization of phonon modes.
A new interesting point is, that FK model does {\it not} require
any external disorder: randomization of bricks order in FK chain occurs
dynamically. Another interesting feature, which characterizes a degree
of chaotization of bricks in CES, is that all the examples of exponential
localization are obtained from {\it single} arbitrary taken CES,
without averaging over any encemble of nearest CES. 

A typical quantity used traditionally in studies of localization phenomena
(see, in particular \cite{Burk96,Tong99}) is the participation ratio
(PR), defined as
\begin{equation}
\label{iprDef}
R=\frac{1}{s}\left( \sum _{i=1}^{s}\psi ^{4}_{i}\right) ^{-1},
\end{equation}
 where $ \psi $ is a normalized ($ \sum _{i}\psi _{i}^{2}=1 $)
vector of the phonon eigenstate. Its value correspond to a chain fraction
occupied by the localized state, but whether the state is localized
exponentially can be unclear. Typical feature of exponentially localized
state is that components outside the center of localization are exponentially
small. Meanwhile, this components do not contribute in PR at all.

To our opinion a better characteristics to indicate an {\it exponential}
localization can be the generalized mean geometrical value (MGV),
defined as\begin{equation}
\label{mgv}
W=s\cdot (\prod _{i=1}^{s}\psi ^{2})^{1/s}=s\exp (\frac{1}{s}\sum ^{s}_{i=1}\ln (\psi ^{2})),
\end{equation}
 where, a normalization factor $ s $ provides that for extended
states $ W $ be order of unity. The reason in favor to $ W $
is that it essentially better probes the exponentially small components
(tails) of the phonon eigenstate. In particular, for a typical exponentially
localized state $ |\psi _{i}|\sim \ell ^{-1/2}\exp (-|i-i_{0}|/\ell ) $,
$ \ell  $ is a localization length, and value of the MGV: $ W\sim \ell ^{-1}\exp (-s/2\ell ) $,
i.e. becomes \emph{exponentially} small. Moreover, one can get from
the value of MGV an estimate of the localization length as
\begin{equation}
\label{Lloc}
\ell =-s/2\ln \left( \frac{W\ell }{s}\right) \simeq -s/2\ln \left( -W/2\ln W\right) .
\end{equation}
Note an important difference between the estimate of the localization
length as inverse participation ratio and our estimate (\ref{Lloc}).
The former estimates a size of the domain where the eigenstate is
localized, and is sensitive to particular short range dynamics for
formation of the given state. On the contrary, the latter is related
to the rate of the exponential falloff at the tails of the eigenstates,
and characterizes properties of the disordered media.

\begin{figure}
{\centering \resizebox*{0.9\columnwidth}{!}{\rotatebox{90}{\includegraphics{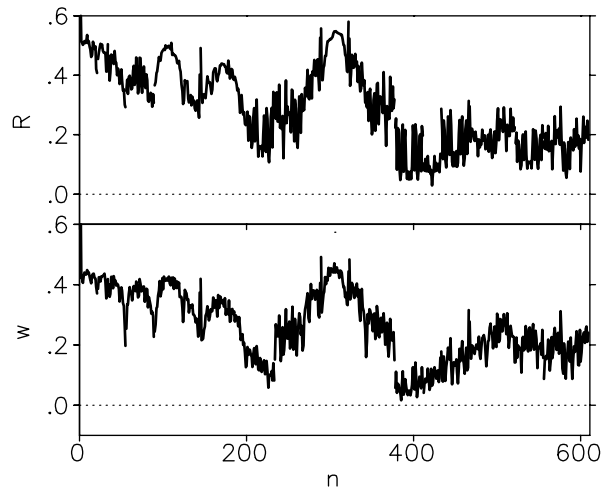}}} \par}

\caption{\label{fGS_PR-MGV}Comparison of PR (upper plot) and MGV (bottom
plot) versus phonon mode number \protect\( n\protect \) for the ground
state of the chain.}
\end{figure}

Now let us turn to localization properties of phonons in the FK chain.
As in the previous section, we concentrate on the numerical study
of the chain with the rotation number $ \nu =r/s=377/610 $, and
parameter $ K=2 $. In Fig.\ref{fGS_PR-MGV} we present our comparison
of PR and MGV for GS. One can see, that both quantities $ R $ and
$ W $ look very similar. Note, that $ W $ for GS is well distinct
from zero (dotted line). This means, that all phonon states are \emph{not}
exponentially localized; the smallness of $ R $ and $ W $ for
some phonon modes means only that this modes are prelocalized only,
which agree with earlier studies \cite{Keto97,Tong99}.

\begin{figure}
{\centering \resizebox*{0.9\columnwidth}{!}{\rotatebox{90}{\includegraphics{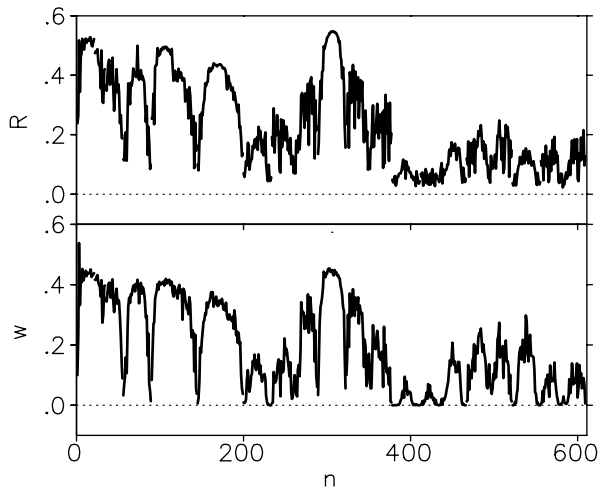}}} \par}

\caption{\label{eLoc}Comparison of PR (upper plot) and MGV (bottom plot)
versus phonon mode number \protect\( n\protect \) for a typical CES
with \protect$ \Delta U=10^{-12}\protect $.}
\end{figure}

In order to see, how the exponential localization manifest itself
in PR and MGV, let us address to CES. We start with a typical CES,
which is still in energy very close to GS: $ \Delta U=U_{CES}-U_{GS}=10^{-12} $.
Despite to, that energy splitting is small, this CES belong to 5th
band in the energy structure of CES \cite{Zh01}, and it is one of
$ 10^{9} $ CESs with $ \Delta U\lesssim 10^{-12} $. In Fig.\ref{eLoc}
we plot $ R $ and $ W $ for the same CES. It is seen that the
behaviour of PR remains in main details the same, while in the MGV
plot there are points, where the curve touches a \emph{zero} line.
This means, that $ W $ at these points is exponentially small,
i.e. there is an \emph{exponential} localization of these modes. Note,
that the PR plot at the same points has no clear indications that
these modes are localized exponentially.

\begin{figure}
{\centering \resizebox*{0.9\columnwidth}{!}{\rotatebox{90}{\includegraphics{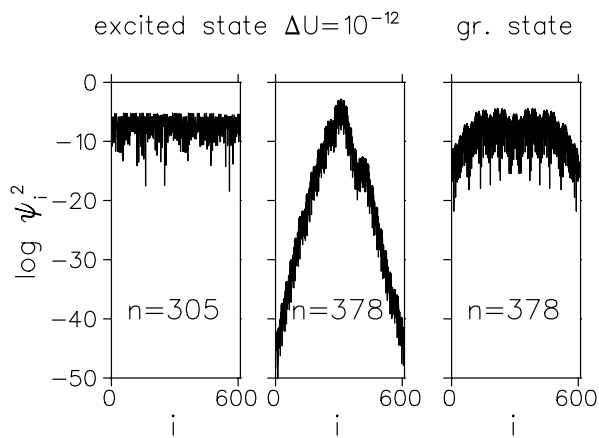}}} \par}

\caption{\label{exMods}Phonon eigenvector component distribution versus the
component number \protect\( i\protect \). Examples of non-localized
and localized modes.}
\end{figure}

Some typical examples of nonlocalized mode \( (n=305) \) and localized
one \( (n=378) \) for this CES is presented in Fig.\ref{exMods}.
We see, that the latter is perfectly localized exponentially, with
a localization length \( \ell \sim 12 \), which characterizes a correlation
length of disorder. For comparison we show also the same mode in the
GS in a ``prelocalized'' state.

\begin{figure}
{\centering \resizebox*{0.9\columnwidth}{!}{\rotatebox{90}{\includegraphics{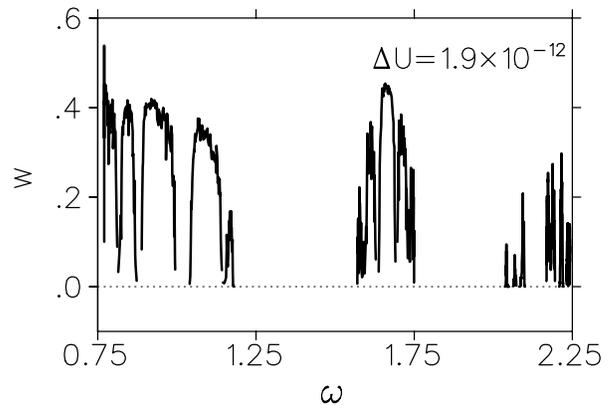}}} \par}

\caption{\label{wFr}MGV versus the frequency \protect\( \omega \protect \)
of the phonon. It is seen, that phonon are exponentially localized
ot the edges of ferquency bands.}
\end{figure}

In fact, CES with \( \Delta U=10^{-12} \) corresponds an early localization
of phonon modes: only small fraction of the is exponentially localized,
as seen from Fig.\ref{eLoc}. To get insight, what modes are localized
at the first turn, in Fig.\ref{wFr} we plot MGV as a function of
phonon frequency. We see, that localized modes are located at edges
of the frequency spectrum bands.

\begin{figure}
{\centering \resizebox*{0.9\columnwidth}{!}{\rotatebox{90}{\includegraphics{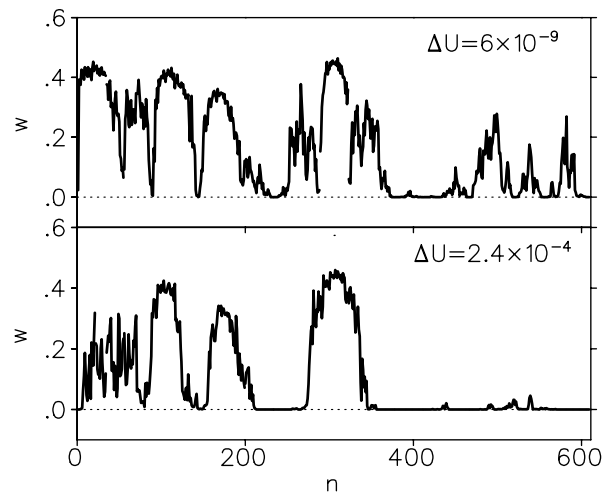}}} \par}

\caption{\label{exW}The MGV versus phonon mode number \protect\( n\protect \)
in two CES.}
\end{figure}

Now, how this picture localization looks for CES with higher splitting
from GS? We expect \cite{Zh01}, that at the splitting \( \Delta U=10^{-12} \)
the largest robust elementary cells\footnote{%
Let us remind (see previous section), that an elementary cell is a
brick with one glue particle added.
} 
are \( 13 \) and \( 21 \), that take part in mutual permutations
only, while larger cells are destroyed. In the range of splitting
\( \Delta U=10^{-9}\div 10^{-8} \) the cell \( 21 \) can dissociate
\cite{Zh01},
that increase a number of smaller cells \( 13 \) and decrease a corelation
length of the disorder. Next, at splitting \( \Delta U=10^{-5}\div 10^{-4} \)
the cell \( 13 \) can dissociate too, and permutations of cells \( 5 \)
come into play, that decrease a correlation length of the disorder
even more. In apparent agreement with our expectations decrease of
the disorder scale results in a total exponential localization of
substantial fraction of phonon modes, especially in high frequency
regoin, as seen from Fig.\ref{exW}. 

\begin{figure}
{\centering \resizebox*{0.9\columnwidth}{!}{\rotatebox{90}{\includegraphics{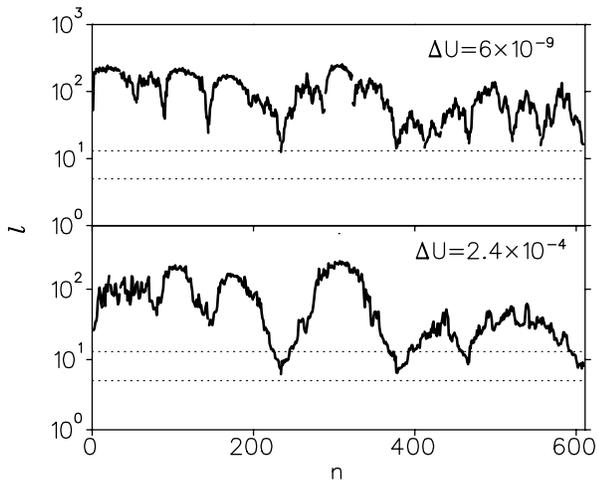}}} \par}

\caption{\label{loc2EX}Localization length \protect\( \ell \protect \) versus
phonon mode number \protect\( n\protect \). Horizontal dotted lines
correspond to levels \protect\( \ell =5\protect \) and \protect\( 13\protect \).}
\end{figure}

In Fig.\ref{loc2EX} we plot fo the same two CES our estimate for
localization length (\ref{Lloc}). Two dotted lines show levels \( \ell =5 \)
and \( 13 \), which correspond to expected sizes of largest robust
elementary cells, which survive at energy splitting \( \Delta U\sim 10^{-4} \)
and \( 10^{-8} \), respectively. We see, that minimal localization
length follows the size of maximal robust structure in the chain.

\begin{figure}
{\centering \resizebox*{0.9\columnwidth}{!}{\rotatebox{90}{\includegraphics{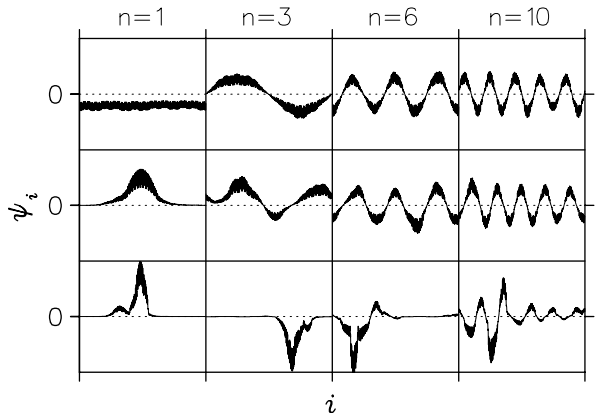}}} \par}

\caption{\label{lwStates}Profiles of phonon eigenstates: eigenvector component
distribution versus component number. The upper corresponds to CS,
while next two present CES with splitting from GS \protect\( \Delta U=6\cdot 10^{-8}\protect \)
and \protect\( 2.4\cdot 10^{-4}\protect \) respectively. }
\end{figure}

Since the chain is not homogeneous, localization properties are not
homogeneous too. As seen from Figs. \ref{exW},\ref{loc2EX}, the
localization is maximal at the high frequency part of phonon spectrum,
while low frequency part seems nonlocalized. In fact, at least longwave
modes of phonons show a clear tendency to be localized too, see Fig.\ref{lwStates}. 

\section{Discussion and conclusions.}

In this paper we have studied properties of phonon modes in the Frenkel-Kontorova
chain, taken in the regime of pinned phase. Spatial "brick" structure of the ground
state of the chain found in \cite{Zh01}, appears to be very useful for 
understanding the fine structure of the chain phonon spectrum. Actually, similar
analysis can also be performed for electronic spectrum, studied in recent work
\cite{Tong02}. 

However, it is obvious, that in a real physical scale the ground state is highly
degenerated with a huge number of the static configurational excitations states (CES)
of the chain. Actually this means that the true ground state of the chain is
practically inaccessible. CES has properties quite different from that of the ground state:
their spatial structure is rather chaotic \cite{Zh01}. As a result,  they can  cause 
the Anderson-like (exponential) localization of phonon modes similar to that seen in
disordered media. This means, that results of previous 
studies of phonon\cite{Burk96,Keto97,Tong99,Hu00} and electron properties\cite{Tong02} 
should be revized or extended to more realistic states of the chain.

Configurational excitation states (CES) can provide a possibility
to study a gradual transition from the order to disorder. It is important,
that the number of CES grows with energy splitting from the GS very
fast. Even at very small splitting the number of ECS is huge. It is curious
that each particular CES has intrinsic chaos, which reminds the situation in
classical spin glass \cite{Mez97}, but in contrast to the latter,
the chaos in ECS arises dynamically, without any external noise. 

The quantization of FK model in small $ \hbar$ limit is in essence
the quantization of its phonon modes \cite{Hu01}. Therefore localization of phonon
modes means localization of quantum states. Note also, that phonon
quantization problem is very close to electron quantization problem \cite{Tong02},
where transition from localization to delocalized state is interesting
as a insulator-metal transition .

This work was supported in part by EC RTN network contract HPRN-CT-2000-0156.
One of us (O.V.Z.) thanks Cariplo fundation, INFN and RFBR grant No.04-02-16570
for financial support. Support from the PA INFN ``Quantum transport
and classical chaos'' is gratefully acknowledged.

\end{document}